\documentclass[conference, a4paper]{IEEEtran}
\usepackage{graphicx}
\usepackage{amsmath}
\usepackage{amsfonts}
\usepackage{amssymb}

\newtheorem{theorem}{Theorem}

\newenvironment{proof}[1][Proof]{\noindent\textbf{#1.} }{\ \rule{0.5em}{0.5em}}
\begin{document}

\title{Optimal Power Allocation for Three-phase Bidirectional DF Relaying
with Fixed Rates }
\author{%
\IEEEauthorblockN{Zoran Hadzi-Velkov\IEEEauthorrefmark{1}, Nikola
Zlatanov\IEEEauthorrefmark{2}, and Robert
Schober\IEEEauthorrefmark{3} \vspace{3mm} }
\IEEEauthorblockA{\IEEEauthorrefmark{1}Faculty of Electrical
Engineering and Information Technologies, Ss. Cyril and Methodius
University, Skopje, Macedonia \vspace{-0.0mm}}
\IEEEauthorblockA{\IEEEauthorrefmark{2}Department of Electrical
and Computer Engineering, University of British Columbia,
Vancouver, Canada
\IEEEauthorblockA{\IEEEauthorrefmark{3}Department of Electrical,
Electronics and Communications Engineering, University of
Erlangen-Nuremberg, Germany \vspace{-0.0mm}} \vspace{-8mm}}}

\maketitle

\let\thefootnote\relax\footnotetext{This work has been supported by the Alexander
von Humboldt fellowship program for experienced researchers}

\begin{abstract}
Wireless systems that carry delay-sensitive information (such as speech
and/or video signals) typically transmit with fixed data rates, but may
occasionally suffer from transmission outages caused by the random nature of
the fading channels. If the transmitter has instantaneous channel state
information (CSI) available, it can compensate for a significant portion of
these outages by utilizing power allocation. In this paper, we consider
optimal power allocation for a conventional dual-hop bidirectional
decode-and-forward (DF) relaying system with a three-phase transmission
protocol. The proposed strategy minimizes the average power consumed by the
end nodes and the relay, subject to some maximum allowable system outage
probability (OP), or equivalently, minimizes the system OP while meeting
average power constraints at the end nodes and the relay. We show that in
the proposed power allocation scheme, the end nodes and the relay adjust
their output powers to the minimum level required to avoid outages, but will
sometimes be silent, in order to conserve power and prolong their lifetimes.
For the proposed scheme, the end nodes use the instantaneous CSI of their
respective source-relay links and the relay uses the instantaneous CSI of
both links.
\end{abstract}

\vspace{-4mm}

\section{Introduction}

Compared to unidirectional relaying, bidirectional (two-way)
relaying is a more suitable alternative for applications where the
end nodes intend to exchange information (e.g., in interacive
applications) \cite{R1}. Bidirectional relaying is also spectrally
more efficient than unidirectional relaying, because it exploits
the broadcast nature of the wireless medium \cite{R2}. The relay
combines two unidirectional unicast transmissions into a single
broadcast transmission using the network coding concept \cite{R3}.
The time-division broadcast (TDBC) protocol is the most important
three-phase bidirectional decode-and-forwad (DF) relaying scheme.
It increases spectral efficiency by 33\% compared to
unidirectional relaying \cite{R4}.

Papers \cite{R5} and \cite{R6} study the capacity outage probability (OP) of
the TDBC protocol. The OP well describes the behaviour of a system that
operates at fixed information rates over quasi-static (i.e., slowly fading)
channels, where the rate is selected such that each codeword is transmitted
over one channel realization. Fixed-rate communication is suitable for
delay-sensitive applications, such as, bidirectional interactive speech
and/or video communication. However, \cite{R5} and \cite{R6} assume that the
end nodes and the relay transmit at fixed output powers, although they have
to have instantaneous channel state information (CSI) available as this is
required for decoding. Apart from decoding, this CSI can also be used for
power adaptation at the relay and the end nodes such that outage
minimization is achieved under some \textit{(long-term) average power}
constraint. Unlike the more restrictive short-term power constraint that
limits the codeword power for each channel realization, the average power
constraints limit the average power of all codewords over all channel
realizations \cite{R10}. For point-to-point channels, such power adaptation
is known as \textit{truncated channel inversion} and has been introduced in
\cite{R9}. For unidirectional relaying, optimal power allocation for source
and relay has been studied for both conventional amplify-and-forward \cite%
{R7} and decode-and-forward (DF) \cite{R8} relaying systems under various
average power constraints. Optimal power allocation has been shown to
introduce significant performance improvement relative to constant power
transmissions \cite{R7}-\cite{R10}. However, literature does not offer
similar results for bidirectional relaying.

In this work, we derive power control strategyies for the end-nodes and the
relay in three-phase bidirectional DF relaying. For predefined constant
rates in both directions, the proposed power allocation achieves
minimization of the system OP assuming individual average power constraints
at the end nodes and the relay. For the power allocation, the end nodes use
the instantaneous CSI of their respective source-relay links and the relay
uses the instantaneous CSI of both links. Intuitively, it is not necessary
for the end nodes and the relay to transmit at their maximum available power
in each transmission cycle, but transmit with the minimum power required to
avoid outages, or sometimes even be silent when an outage is unavoidable,
thus conserving their power. In other words, we allow outages to occur in
cases of deep fades, but for the rest of the time we ensure successful
transmissions at the predefined constant transmission rate. \ \ \ \

\vspace{-1mm}

\section{System and channel model}

The considered bidirectional relaying system consists of two end-nodes ($%
S_{1}$ and $S_{2}$) and a half-duplex DF relay $R$. The bidirectional
communication consists of two parallel unidirectional communication
sessions, $S_{1}\rightarrow S_{2}$ and $S_{2}\rightarrow S_{1}$. Each
communication session is realized at a fixed information rate, $R_{01}$ and $%
R_{02}$, respectively. We assume the direct $S_{1}-S_{2}$ link is not
available, thus, the bidirectional communication is realized only via the
relayed link. The OP for this system is defined as the probability that at
least one (or both) of the communications sessions is in outage.

The squared amplitudes of the $S_{1}-R$ and $S_{2}-R$ channels are denoted
by $x$ and $y$, and have arbitrary average values $\Omega _{X}$ and $\Omega
_{Y}$, respectively. We assume that the $R-S_{1}$ channel is reciprocal to
the $S_{1}-R$ channel, and the $R-S_{2}$ channel is reciprocal to the $%
S_{2}-R$ channel. We adopt the Rayleigh block fading model, which means that
the values of $x$ and $y$ are constant within each transmission cycle, but
change from one transmission cycle to the next. Thus, in each transmission
cycle, the pair $\left( x,y\right) $ denotes the channel state, where both $%
x $ and $y$ follow the Rayleigh probability distribution function (PDF).

Each transmission cycle is divided into three phases: In phase 1, $S_{1}$%
transmits its codeword $s_{1}(t)$ at information rate $R_{01}$ with an
output power $P_{S1}\left( x\right) $, and the DF relay receives. In phase
2, $S_{2}$ transmits its codeword $s_{2}(t)$ at information rate $R_{02}$
with an output power $P_{S2}\left( y\right) $, and the DF relay receives.
The relay attempts to decode both codewords $s_{1}(t)$ and $s_{2}(t)$. If it
successfully decodes both of them, the relay generates a single composite
codeword that carries the information of both $s_{1}(t)$ and $s_{2}(t)$ \cite%
{R2}. Then, in phase 3, the relay broadcasts the composite codeword towards
the end nodes with an output power $P_{R}\left( x,y\right) $.

Note that the output powers from the end nodes and the relay have been
written as functions of the channel state. In each transmission cycle, node $%
S_{1}$ is assumed to know only the value of $x$, whereas node $S_{2}$ is
assumed to know only the value of $y$. Based on this CSI knowledge, each end
node can "subtract" its own codeword from the received codeword, and then
attempt to decode the noisy version of the codeword that originates from the
other node. The DF relay can decode both codewords since it is assumed to
know both $x$ and $y$. The received signals at the end nodes and the relay
are corrupted by additive white Gaussian noise (AWGN) with zero mean and
unit variance.

We present optimal power allocation (OPA) strategies at the end nodes and
the relay that minimize the system OP, subject to the individual average
power constraints $\overline{P}_{S1}$, $\overline{P}_{S2}$, and $P_{avg}$ at
$S_{1}$, $S_{2}$, and the relay. In phase 1, $S_{1}$ uses power control
based on its knowledge of $x$ so as to achieve truncated channel inversion
of the $S_{1}-R$ channel at rate $R_{01}$ \cite{R9}. Similarly, in phase 2, $%
S_{2}$ uses power control based on its knowledge of $y$ so as to achieve
truncated channel inversion of the $S_{2}-R$ channel at rate $R_{02}$. In
phase 3, if both codewords $s_{1}(t)$ and $s_{2}(t)$ are successfully
decoded by the relay, the adopted power control\ mechanism minimizes the OP
over the broadcast channel (Section IV).

\section{Power control at the end nodes}

The $S_{1}-R$ channel can support $S_{1}$'s transmission rate, $R_{01}$, if
the instantaneous capacity of the channel between $S_{1}$ and the relay
exceeds this rate,
\begin{equation}
\frac{1}{3}\log (1+P_{S1}x)\geq R_{01}\text{,}  \label{Eq1}
\end{equation}%
where the the pre-log factor $1/3$ is due to the three-phase transmission
cycle. For an available average power of end node $S_{1}$ of $\overline{P}%
_{S1}$, the OPA strategy at $S_{1}$ is given by \cite{R8}
\begin{equation}
P_{S1}(x)=\left\{
\begin{array}{c}
\frac{\delta _{1}}{x},\text{ \ }x\geq x_{0} \\
0,\text{ \ }x<x_{0}%
\end{array}%
\right. \text{,}  \label{Eq2}
\end{equation}%
where $\delta _{1}=2^{3R_{01}}-1$. In (\ref{Eq2}), the cutoff threshold $%
x_{0}$ is determined from
\begin{equation}
\overline{P}_{S1}=\int_{x_{0}}^{\infty }\frac{\delta _{1}}{x}f_{X}\left(
x\right) dx=\frac{\delta _{1}}{\Omega _{X}}E_{1}\left( \frac{x_{0}}{\Omega
_{X}}\right) \text{.}  \label{Eq3}
\end{equation}%
The right hand side of (\ref{Eq3}) is valid for Rayleigh fading, where $%
E_{1}\left( \cdot \right) $ is the exponential integral function \cite{R11}.
Analogously, the $S_{2}-R$ channel can support $S_{2}$'s transmission rate, $%
R_{02}$, with its average output power $\overline{P}_{S2}$, if the
instantaneous capacity of the channel between $S_{2}$ and the relay exceeds
this rate,
\begin{equation}
\frac{1}{3}\log (1+P_{S2}y)\geq R_{02}  \label{Eq4}
\end{equation}%
which leads to the OPA strategy at $S_{2}$,
\begin{equation}
P_{S2}(y)=\left\{
\begin{array}{c}
\frac{\delta _{2}}{y},\text{ \ }y\geq y_{0} \\
0,\text{ \ }y<y_{0}%
\end{array}%
\right. \text{,}  \label{Eq5}
\end{equation}%
where $\delta _{2}=2^{3R_{02}}-1$. The corresponding cutoff threshold $y_{0}$
is determined from
\begin{equation}
\overline{P}_{S2}=\int_{y_{0}}^{\infty }\frac{\delta _{2}}{y}f_{Y}\left(
y\right) dy=\frac{\delta _{2}}{\Omega _{Y}}E_{1}\left( \frac{y_{0}}{\Omega
_{Y}}\right) \text{.}  \label{Eq6}
\end{equation}

\section{Power control at the relay}
Considering (\ref{Eq2}) and (\ref{Eq5}), the relay successfully decodes both
codewords $s_{1}(t)$ and $s_{2}(t)$ only if $\left( x,y\right) \in D_{R}$,
where $D_{R}$ is the relay's non-outage region defined as
\begin{equation}
D_{R}\text{: \ \ \ \ }x\geq x_{0}\text{ and }y\geq y_{0}.  \label{Eq7}
\end{equation}%
In this case, the relay generates the composite signal, and, in phase 3
broadcasts the composite codeword towards the end nodes with an output power
$P_{R}\left( x,y\right) $.

\begin{theorem}
The solution of the optimization problem%
\begin{equation*}
\text{$\underset{P_{R}\left( x,y\right) }{\text{minimize}}$ }P_{R}\left(
x,y\right)
\end{equation*}%
\vspace{-5mm}
\begin{eqnarray}
\text{subject to \ \ }\frac{1}{3}\log _{2}\left( 1+P_{R}\left( x,y\right)
\text{ }y\right)  &\geq &R_{01}  \notag \\
\frac{1}{3}\log {}_{2}\left( 1+P_{R}\left( x,y\right) \text{ }x\right)
&\geq &R_{02}  \notag \\
x \geq x_{0}\text{ and }y\geq y_{0}  \label{7a}
\end{eqnarray}
\end{theorem}

is given by
\begin{equation}
P_{R,st}\left( x,y\right) =\left\{
\begin{array}{c}
\max \left\{ \frac{\delta _{1}}{y},\frac{\delta _{2}}{x}\right\} ,\text{ \ }%
x\geq x_{0}\text{ and }y\geq y_{0} \\
0,\text{ otherwise}%
\end{array}%
\right. .  \label{10}
\end{equation}

\begin{proof}
Optimization problem (\ref{7a}) is a standard \textit{linear programming}
(LP) problem, whose solution is feasible because the intersection of the
constraints is a non-empty set, and it lies at the boundary of the
intersection, given by (\ref{10}). This concludes the proof.
\end{proof}

If the relay successfully decodes both codewords, i.e., $\left( x,y\right)
\in D_{R}$, the power adaptation (\ref{10}) guarantees zero outages in the
broadcase phase with minimum output power from the relay, because both end
nodes can successfully decode their intended codewords in the broadcast
phase. If $\left( x,y\right) \notin D_{R}$, the relay should be silent as it
cannot decode at least one of the two codewords. Assuming power control
accoding to (\ref{10}), the relay's non-outage region $D_{R}$ is divided
into two non-overlaping regions, $D_{R}=D_{R}^{\prime }\cup D_{R}^{\prime
\prime }$, such that
\begin{equation}
D_{R}^{\prime }:\text{ \ \ }x\geq x_{0}\text{ and }y_{0}\leq y\leq \frac{%
\delta _{1}}{\delta _{2}}x  \label{13a}
\end{equation}%
\vspace{-2mm}
\begin{equation}
D_{R}^{\prime \prime }:\text{ \ \ }y\geq y_{0}\text{ and }x_{0}\leq x\leq
\frac{\delta _{2}}{\delta _{1}}y  \label{13b}
\end{equation}%
If $\left( x,y\right) \in D_{R}^{\prime }$, the relay transmits with power $%
\delta _{1}/y$, whereas if $\left( x,y\right) \in D_{R}^{\prime \prime }$,
the relay transmits with power $\delta _{2}/x$. Figs. 1 and 2 graphically
illustrate the non-outage regions $D_{R}^{\prime }$ given by (\ref{13a}) and
$D_{R}^{\prime \prime }$ given by (\ref{13b}), where Fig. 1 applies to the
case $\delta _{2}y_{0}\leq \delta _{1}x_{0}$ whereas Fig. 2 applies to the
case $\delta _{2}y_{0}>\delta _{1}x_{0}$. Thus, the power adaptation (\ref%
{10}) can be further decomposed as
\begin{equation}
P_{R,st}\left( x,y\right) =\left\{
\begin{array}{c}
\frac{\delta _{1}}{y},\text{ \ \ }x\geq x_{0}\text{ and }y_{0}\leq y\leq
\frac{\delta _{1}}{\delta _{2}}x \\
\frac{\delta _{2}}{x},\text{ \ \ }y\geq y_{0}\text{ and }x_{0}\leq x\leq
\frac{\delta _{2}}{\delta _{1}}y \\
0,\text{ \ \ otherwise}%
\end{array}%
\right. \text{ \ .}  \label{10a}
\end{equation}

\begin{figure}[tbp]
\centering
\includegraphics[width=3.0in]{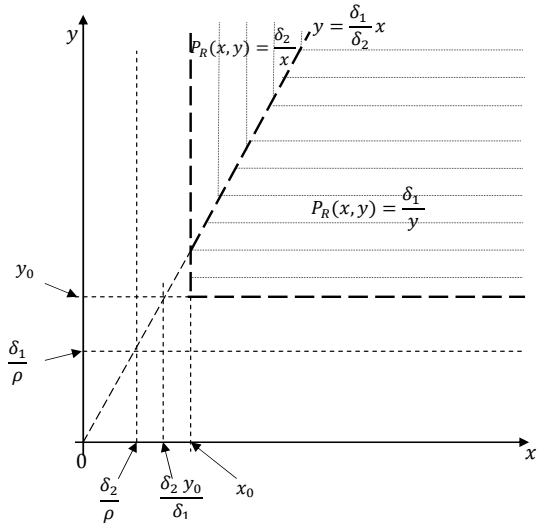} \vspace{-5mm}
\caption{Non-outage region for $\protect\delta_{2}y_{0} \leq \protect\delta%
_{1}x_{0}$} \vspace{-4mm}
\label{fig1}
\end{figure}

\subsection{Outage Minimization}
However, even if $\left( x,y\right) \in D_{R}$, it may still be impossible
to maintain zero outage probability in the broadcast phase, because the
relay is also constrained by its own long-term power budget $P_{avg}$. The
system OP\ is determined as
\begin{eqnarray}
P_{out} =\underset{\left( x,y\right) \in D_{R}}{\int \int }dxdy\text{ }%
f_{X}\left( x\right) f_{Y}\left( y\right) \qquad \qquad \qquad \qquad \qquad
\quad   \notag \\
\times \Pr \left\{ \left.
\begin{array}{c}
\frac{1}{3}\log _{2}\left( 1+P_{R}\left( x,y\right) \text{ }y\leq
R_{01}\right) \text{ \ } \\
\text{OR \thinspace }\frac{1}{3}\log _{2}\left( 1+P_{R}\left( x,y\right)
\text{ }x\leq R_{02}\right)
\end{array}%
\right\vert \left( x,y\right) \in D_{R}\right\} \text{.}  \notag  \label{OP}
\end{eqnarray}%
\begin{equation}
\end{equation}%
According to (\ref{OP}), $P_{out}$ can be minimized if both of the following
requirements are satisfied: (\textit{i}) the area of $D_{R}$ is maximized,
and (\textit{ii}) the conditional outage probability in the integrand of (%
\ref{OP}) is minimized. For the CSI available at the end nodes ($S_{1}$
knows only $x$ and $S_{2}$ knows only $y$), (\ref{Eq3}) and (\ref{Eq6})
guarantee that the area of $D_{R}$ is maximized. For requirement (\textit{ii}%
), we have the following theorem.

\begin{theorem}
The solution of optimization problem
\begin{equation*}
\text{$\underset{P_{R}\left( x,y\right) }{\text{minimize}}$ } \Pr \left\{
\begin{array}{c}
\text{ \ \ \ \ \ \ }\frac{1}{3}\log _{2}\left( 1+P_{R}\left( x,y\right)
\text{ }y\right) \leq R_{01}\text{ } \\
\text{OR \ }\frac{1}{3}\log _{2}\left( 1+P_{R}\left( x,y\right) \text{ }%
x\right) \leq R_{02}%
\end{array}%
\right\}
\end{equation*}%
\begin{equation}
\text{subject to \ \ \ }E_{XY}\left[ P_{R}\left( x,y\right) \right] \leq
P_{avg}\text{ \ and \ }\left( x,y\right) \in D_{R}\text{,}  \label{15}
\end{equation}%
where $E_{XY}\left[ \cdot \right] $ denotes expectation with respect to $X$
and $Y$, is given by
\end{theorem}
\begin{equation}
P_{R}^{\ast }\left( x,y\right) =\left\{
\begin{array}{c}
P_{R,st}\left( x,y\right) ,\text{ \ \ if }P_{R,st}\left( x,y\right) \leq \rho
\\
0,\text{ \ \ \ \ \ \ \ \ \ \ \ \ \ \ \ if }P_{R,st}\left( x,y\right) >\rho%
\end{array}%
\right. \text{ ,}  \label{14}
\end{equation}%
where $P_{R,st}\left( x,y\right) $ is given by (\ref{10a}) and the cutoff
threshold $\rho $ is determined from
\begin{equation}
P_{avg}=E_{XY}\left[ P_{R,st}\left( x,y\right) \text{ }|P_{R,st}\left(
x,y\right) \leq \rho \right] \text{ .}  \label{14a}
\end{equation}

\begin{proof}
The proof is analogous to [8, Appendix B] and [10, Appendix D], where $%
P_{R,st}\left( x,y\right) $ is the minimum short-term power of the relay
that maintains zero outages in the broadcast phase.
\end{proof}

\begin{figure}[tbp]
\centering
\includegraphics[width=3.3in]{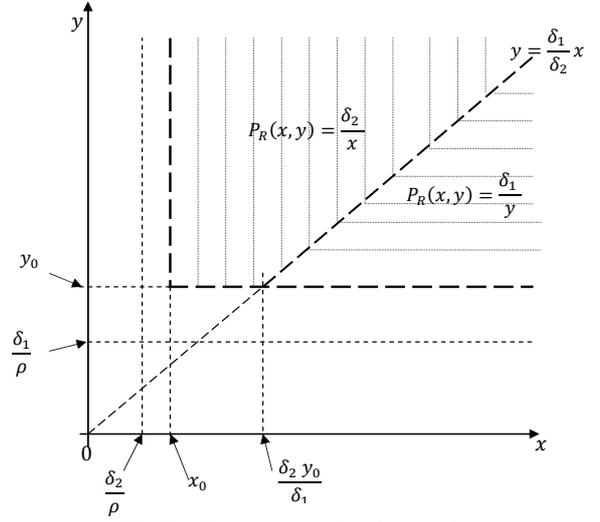} \vspace{-5mm}
\caption{Non-outage region $\protect\delta_{2}y_{0} > \protect\delta%
_{1}x_{0} $} \vspace{-4mm}
\label{fig2}
\end{figure}

Combining (\ref{14}) and (\ref{10a}), the OPA strategy at the relay is
finally obtained as
\begin{equation}
P_{R}^{\ast }\left( x,y\right) =\left\{
\begin{array}{c}
\frac{\delta _{1}}{y},\text{ \ }x\geq x_{0}\text{ and }\max \left\{ \frac{%
\delta _{1}}{\rho },y_{0}\right\} \leq y\leq \frac{\delta _{1}}{\delta _{2}}x
\\
\frac{\delta _{2}}{x},\text{ \ }y\geq y_{0}\text{ and }\max \left\{ \frac{%
\delta _{2}}{\rho },x_{0}\right\} \leq x\leq \frac{\delta _{2}}{\delta _{1}}y
\\
0,\text{ otherwise}%
\end{array}%
\right. \text{ .}  \label{14b}
\end{equation}

Considering the assumptions for the CSI\ availability at the end nodes and
the relay, the power allocation rules $P_{S1}\left( x\right) $, $%
P_{S2}\left( y\right) $, and $P_{R}\left( x,y\right) $, given by (\ref{Eq3}%
), (\ref{Eq5}), and (\ref{14b}), respectively, are optimal as they minimize
the OP of the three-phase bidirectional DF relaying system for individual
power constraints at the end nodes and the relay.

\subsection{Average Output Power and Cutoff Threshold $\protect\rho $}

The analytical expression for the average relay output power is obtained by
statistical averaging of (\ref{14b}) with respect to $x$ and $y$. Let us
first introduce two auxiliary variables $\lambda _{1}$ and $\lambda _{2}$,
defined as
\begin{eqnarray}
\lambda _{1} &=&\max \left\{ x_{0},\delta _{2}/\rho \right\} \text{, }
\notag \\
\lambda _{2} &=&\max \left\{ y_{0},\delta _{1}/\rho \right\} \text{.}
\label{31}
\end{eqnarray}%
a) If $\delta _{2}y_{0}\leq \delta _{1}x_{0}$ (Fig. 1), the average relay
output power, $\overline{P}_{R1}$, is expressed as
\begin{eqnarray}
\overline{P}_{R1}\left( \rho \right) &=&\int\limits_{x_{0}}^{\infty
}f_{X}\left( x\right) \int\limits_{\lambda _{2}}^{\frac{\delta _{1}x}{\delta
_{2}}}\frac{\delta _{1}}{y}f_{Y}\left( y\right) dydx  \notag \\
&&+\int\limits_{\frac{\delta _{1}\lambda _{1}}{\delta _{2}}}^{\infty
}f_{Y}\left( y\right) \int\limits_{\lambda _{1}}^{\frac{\delta _{2}y}{\delta
_{1}}}\frac{\delta _{2}}{x}f_{X}\left( x\right) dxdy  \notag \\
&=&\frac{\delta _{1}}{\Omega _{Y}}E_{1}\left( \frac{\lambda _{2}}{\Omega _{Y}%
}\right) e^{-\frac{x_{0}}{\Omega _{X}}}-\frac{\delta _{1}}{\Omega _{Y}}%
E_{1}\left( \frac{\delta _{1}x_{0}}{\delta _{2}\Omega _{Y}}\right) e^{-\frac{%
x_{0}}{\Omega _{X}}}  \notag \\
&&+\frac{\delta _{1}}{\Omega _{Y}}E_{1}\left( \left( \frac{1}{\Omega _{X}}+%
\frac{\delta _{1}}{\delta _{2}\Omega _{Y}}\right) x_{0}\right)  \notag \\
&&+\frac{\delta _{2}}{\Omega _{X}}E_{1}\left( \left( \frac{1}{\Omega _{X}}+%
\frac{\delta _{1}}{\delta _{2}\Omega _{Y}}\right) \lambda _{1}\right) \text{.%
}  \label{32}
\end{eqnarray}%
b) If $\delta _{2}y_{0}>\delta _{1}x_{0}$ (Fig. 2), the average relay output
power, $\overline{P}_{R2}$, is expressed as
\begin{eqnarray}
\overline{P}_{R2}\left( \rho \right) &=&\int\limits_{\frac{\delta
_{2}\lambda _{2}}{\delta _{1}}}^{\infty }f_{X}\left( x\right)
\int\limits_{\lambda _{2}}^{\frac{\delta _{1}x}{\delta _{2}}}\frac{\delta
_{1}}{y}f_{Y}\left( y\right) dydx  \notag \\
&&+\int\limits_{y_{0}}^{\infty }f_{Y}\left( y\right) \int\limits_{\lambda
_{1}}^{\frac{\delta _{2}y}{\delta _{1}}}\frac{\delta _{2}}{x}f_{X}\left(
x\right) dxdy  \notag \\
&=&\frac{\delta _{2}}{\Omega _{X}}E_{1}\left( \frac{\lambda _{1}}{\Omega _{X}%
}\right) e^{-\frac{y_{0}}{\Omega _{Y}}}-\frac{\delta _{2}}{\Omega _{X}}%
E_{1}\left( \frac{\delta _{2}y_{0}}{\delta _{1}\Omega _{X}}\right) e^{-\frac{%
y_{0}}{\Omega _{Y}}}  \notag \\
&&+\frac{\delta _{2}}{\Omega _{X}}E_{1}\left( \left( \frac{1}{\Omega _{Y}}+%
\frac{\delta _{2}}{\delta _{1}\Omega _{X}}\right) y_{0}\right)  \notag \\
&&+\frac{\delta _{1}}{\Omega _{Y}}E_{1}\left( \left( \frac{1}{\Omega _{Y}}+%
\frac{\delta _{2}}{\delta _{1}\Omega _{X}}\right) \lambda _{2}\right) \text{.%
}  \label{33}
\end{eqnarray}%
Thus, the cutoff threshold $\rho $ is determined from
\begin{equation}
P_{avg}=\left\{
\begin{array}{c}
\overline{P}_{R1}\left( \rho \right) ,\text{ \ \ if \ }\delta _{2}y_{0}\leq
\delta _{1}x_{0} \\
\overline{P}_{R2}\left( \rho \right) ,\text{ \ \ if \ }\delta
_{2}y_{0}>\delta _{1}x_{0}%
\end{array}%
\right. ,  \label{34}
\end{equation}%
where $\overline{P}_{R1}\left( \rho \right) $ and $\overline{P}_{R2}\left(
\rho \right) $ are given by (\ref{32}) and (\ref{33}), respectively. We note
however that the average relay output power has a maximum $\overline{P}%
_{R}^{\max }$, given by
\begin{equation}
\overline{P}_{R}^{\max }=\left\{
\begin{array}{c}
\overline{P}_{R1}^{\max },\text{ \ \ if \ }\delta _{2}y_{0}\leq \delta
_{1}x_{0} \\
\overline{P}_{R2}^{\max },\text{ \ \ if \ }\delta _{2}y_{0}>\delta _{1}x_{0}%
\end{array}%
\right. .  \label{34a}
\end{equation}%
where
\begin{eqnarray}
\overline{P}_{R1}^{\max } &=&\frac{\delta _{1}}{\Omega _{Y}}e^{-\frac{x_{0}}{%
\Omega _{X}}}\left[ E_{1}\left( \frac{y_{0}}{\Omega _{Y}}\right)
-E_{1}\left( \frac{\delta _{1}x_{0}}{\delta _{2}\Omega _{Y}}\right) \right]
\notag \\
&&+\left( \frac{\delta _{1}}{\Omega _{Y}}+\frac{\delta _{2}}{\Omega _{X}}%
\right) E_{1}\left( \frac{x_{0}}{\Omega _{X}}+\frac{\delta _{1}x_{0}}{\delta
_{2}\Omega _{Y}}\right) \text{,}  \label{35}
\end{eqnarray}%
and
\begin{eqnarray}
\overline{P}_{R2}^{\max } &=&\frac{\delta _{2}}{\Omega _{X}}e^{-\frac{y_{0}}{%
\Omega _{Y}}}\left[ E_{1}\left( \frac{x_{0}}{\Omega _{X}}\right)
-E_{1}\left( \frac{\delta _{2}y_{0}}{\delta _{1}\Omega _{X}}\right) \right]
\notag \\
&&+\left( \frac{\delta _{1}}{\Omega _{Y}}+\frac{\delta _{2}}{\Omega _{X}}%
\right) E_{1}\left( \frac{y_{0}}{\Omega _{Y}}+\frac{\delta _{2}y_{0}}{\delta
_{1}\Omega _{X}}\right)  \label{36}
\end{eqnarray}%
Eq. (\ref{35}) is obtained by setting $\lambda _{1}=x_{0}$ in (\ref{32}),
and (\ref{36}) is obtained by setting $\lambda _{2}=y_{0}$ in (\ref{33})

\subsection{Outage Probability}

From (\ref{14a}) we see that $P_{avg}$ imposes the cutoff threshold $\rho $
that maximizes the non-outage region in the broadcast phase, $D_{B}=\left\{
\left( x,y\right) \left\vert P_{R,st}\left( x,y\right) \leq \rho \right.
\right\} $, leading to the maximization of the system's non-outage region, $%
D=\left( D_{R}^{\prime }\cap D_{B}\right) \cup \left( D_{R}^{\prime \prime
}\cap D_{B}\right) $. As a result, OP\ is minimized and calculated as $%
P_{out}=1-\Pr \left\{ \left( x,y\right) \in D\right\} $.

a) If $\delta _{2}y_{0}\leq \delta _{1}x_{0}$ (Fig. 1), the system OP is
expressed as
\begin{eqnarray}
P_{out,1} &=&1-\int\limits_{x_{0}}^{\infty }f_{X}\left( x\right)
\int\limits_{\lambda _{2}}^{\frac{\delta _{1}x}{\delta _{2}}}f_{Y}\left(
y\right) dydx  \notag \\
&&-\int\limits_{\frac{\delta _{1}\lambda _{1}}{\delta _{2}}}^{\infty
}f_{Y}\left( y\right) \int\limits_{\lambda _{1}}^{\frac{\delta _{2}y}{\delta
_{1}}}f_{X}\left( x\right) dxdy  \notag \\
&=&1-e^{-\frac{x_{0}}{\Omega _{X}}}e^{-\frac{\lambda _{2}}{\Omega _{Y}}%
}-e^{-\lambda _{1}\left( \frac{1}{\Omega _{X}}+\frac{\delta _{1}}{\delta
_{2}\Omega _{Y}}\right) }  \notag \\
&&+\frac{\delta _{2}\Omega _{Y}}{\delta _{1}\Omega _{X}+\delta _{2}\Omega
_{Y}}e^{-x_{0}\left( \frac{1}{\Omega _{X}}+\frac{\delta _{1}}{\delta
_{2}\Omega _{Y}}\right) }  \notag \\
&&+\frac{\delta _{1}\Omega _{X}}{\delta _{1}\Omega _{X}+\delta _{2}\Omega
_{Y}}e^{-\lambda _{1}\left( \frac{1}{\Omega _{X}}+\frac{\delta _{1}}{\delta
_{2}\Omega _{Y}}\right) }\text{.}  \label{37}
\end{eqnarray}%
b) If $\delta _{2}y_{0}>\delta _{1}x_{0}$ (Fig. 2), the system OP is
expressed as
\begin{eqnarray}
P_{out,2} &=&1-\int\limits_{\frac{\delta _{2}\lambda _{2}}{\delta _{1}}%
}^{\infty }f_{X}\left( x\right) \int\limits_{\lambda _{2}}^{\frac{\delta
_{1}x}{\delta _{2}}}f_{Y}\left( y\right) dydx  \notag \\
&&-\int\limits_{y_{0}}^{\infty }f_{Y}\left( y\right) \int\limits_{\lambda
_{1}}^{\frac{\delta _{2}y}{\delta _{1}}}f_{X}\left( x\right) dxdy  \notag \\
&=&1-e^{-\frac{y_{0}}{\Omega _{Y}}}e^{-\frac{\lambda _{1}}{\Omega _{X}}%
}-e^{-\lambda _{2}\left( \frac{1}{\Omega _{Y}}+\frac{\delta _{2}}{\delta
_{1}\Omega _{X}}\right) }  \notag \\
&&+\frac{\delta _{1}\Omega _{X}}{\delta _{1}\Omega _{X}+\delta _{2}\Omega
_{Y}}e^{-y_{0}\left( \frac{1}{\Omega _{Y}}+\frac{\delta _{2}}{\delta
_{1}\Omega _{X}}\right) }  \notag \\
&&+\frac{\delta _{2}\Omega _{Y}}{\delta _{1}\Omega _{X}+\delta _{2}\Omega
_{Y}}e^{-\lambda _{2}\left( \frac{1}{\Omega _{Y}}+\frac{\delta _{2}}{\delta
_{1}\Omega _{X}}\right) }\text{ .}  \label{38}
\end{eqnarray}

\textit{Remark}: When $P_{avg}=\overline{P}_{R}^{\max }$, the OPA strategy
at the relay (\ref{14b}) is transformed to
\begin{equation}
P_{R}^{\ast }\left( x,y\right) =P_{R,st}\left( x,y\right)   \label{16}
\end{equation}%
where $P_{R,st}\left( x,y\right) $ is given by (\ref{10a}), which does not
involve the threshold $\rho $. In this case, the system OP attains its
minimum value given by
\begin{equation}
P_{out}^{\min }=1-e^{-\frac{x_{0}}{\Omega _{x}}}e^{-\frac{y_{0}}{\Omega _{Y}}%
}\text{. }  \label{17}
\end{equation}

\section{Numerical Examples}

In the following two scenarios, we illustrate the performance improvement of
the considered three-node TDBC relaying system with the proposed OPA,
relative to a respective relaying system with fixed power allocation (FPA), $%
P_{S1}^{\text{fix}}$, $P_{S2}^{\text{fix}}$, and $P_{R}^{\text{fix}}$. The
rates are fixed to $R_{01}=R_{02}=1/3$.

\textbf{Scenario 1}: For a given total available power $P_{T}$, the system
with OPA assumes $\overline{P}_{S1}=\overline{P}_{S2}=\overline{P}%
_{R}=P_{T}/3$, whereas the system with FPA assumes $P_{S1}^{\text{fix}%
}=P_{S2}^{\text{fix}}=P_{R}^{\text{fix}}=P_{T}/3$. Fig. 3 shows significant
OP improvement due to the proposed OPA. In each coding block, OPA scheme
allocates just enough power to the end nodes and the relay so as to maintain
the desired rate, and some or all of the nodes are silent when "deep fades"
occur. On the other hand, FPA always spends the same power in each coding
block regardless of the channel state.

\textbf{Scenario 2}: We consider the power gains at the end nodes and the
relay utilizing OPA. The power gain at the end node is defined as $P_{S}^{%
\text{fix}}/\overline{P}_{S}$, whereas the power gain at the relay is
defined as $P_{R}^{\text{fix}}/\overline{P}_{R}^{\max }$. To achieve the
minimum possible OP, the system with OPA assumes $\overline{P}_{S1}=%
\overline{P}_{S2}=\overline{P}_{S}$ and $\overline{P}_{R}=\overline{P}%
_{R}^{\max }$, whereas the system with FPA assumes $P_{S1}^{\text{fix}%
}=P_{S2}^{\text{fix}}=P_{R}^{\text{fix}}$. For a given OP, $\overline{P}_{S}$
is determined from (\ref{17}), whereas, depending on $x_{0}$ and $y_{0}$, $%
\overline{P}_{R}^{\max }$ is determined either from (\ref{35}) or (\ref{36}%
). The minimum OP of the system with FPA is also determined from (\ref{17}),
such that $x_{0}$ is substituted by $\delta _{1}/P_{S1}^{\text{fix}}$ and $%
y_{0}$ is substituted by $\delta _{2}/P_{S2}^{\text{fix}}$, and is achieved
for $P_{R}^{\text{fix}}\geq \max \left\{ \delta _{1}P_{S2}^{\text{fix}%
}/\delta _{2},\delta _{2}P_{S1}^{\text{fix}}/\delta _{1}\right\} $. We set $%
P_{R}^{\text{fix}}$ to its minimum value. According to Fig. 4, the power
gains are remarkably high when the OP is low, because channel inversion is
applied to almost all channel states ($x_{0}$ and $y_{0}$ have low values,
and $\rho $ has high value). For relatively high OPs (OP between 0.3 and
0.7), the power gain is minimized (yet although still above 5 dB), because
the nodes are often silent although the channel states are not exposed to
"deep fades".

\section{Conclusion}

In this paper, we show that the nodes in a bidirectional relaying system can
utilize their available CSI for power control and thus achieve remarkable
performance improvements and/or power savings. The proposed power allocation
strategies at the end nodes and the relay minimize the OP of a conventional
three-phase bidirectional DF relaying system, subject to the long-term
available power budgets at the respective nodes. These benefits come without
additional cost for the system, because the CSI at the end nodes and the
relay have to be available for decoding purposes anyways.

\begin{figure}[tbp]
\centering
\includegraphics[width=3.5in]{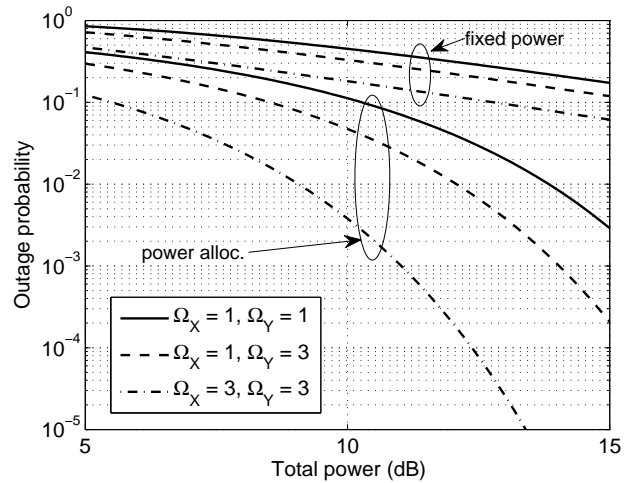} \vspace{-7mm}
\caption{Outage performance improvement due to power allocation} \vspace{-4mm}
\label{fig3}
\end{figure}

\begin{figure}[tbp]
\centering
\includegraphics[width=3.5in]{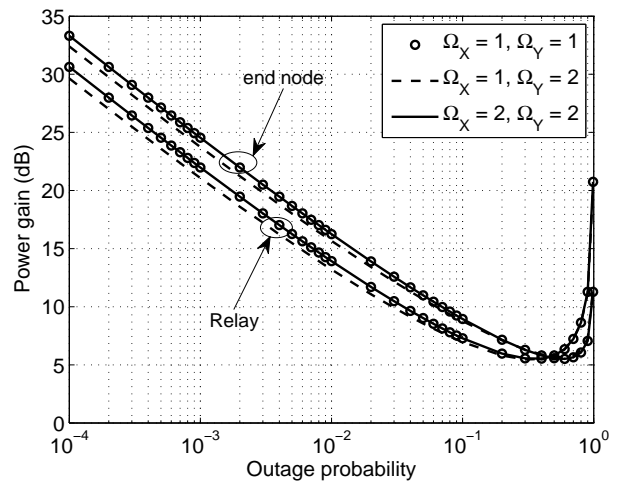} \vspace{-7mm}
\caption{Power savings at the end nodes and the relay} \vspace{-4mm}
\label{fig4}
\end{figure}

\end{document}